
\magnification=\magstep1
\def\ltwid{\mathrel{\raise.3ex\hbox{$<$\kern-.75em\lower1ex\hbox{$\sim$}}}}

\rightline{UFIFT-HEP-95-9}
\bigskip
\bigskip
\bigskip

\centerline {{\bf THE POOLTABLE ANALOGY TO AXION PHYSICS}
\footnote{*} {Based upon a talk at the XXXth Rencontres de Moriond
{\it Dark Matter in Cosmology} and {\it Clocks and Tests of Fundamental
Laws}, Villars-sur-Ollon, Switzerland, Sept. 21-28, 1995.}}
\bigskip
\bigskip
\bigskip
\centerline {\bf Pierre Sikivie}
\bigskip
\centerline {{\it Department of Physics}}
\centerline {{\it University of Florida}}
\centerline {{\it Gainesville, FL 32611}}

\bigskip
\bigskip
\bigskip

\centerline {\bf Abstract}
\bigskip
\bigskip

An imaginary character named TSP finds himself in a playroom
whose floor is tilted to one side.  However, the pooltable in the
playroom is horizontal.  TSP wonders how this can be.  In doing so,
he embarks upon an intellectual journey which parallels that which
has been travelled during the past two decades by physicists
interested in the Strong $CP$ Problem and axion physics.

\vfill\eject

        Consider the physics involved in playing snooker.  The rules
of the game require that the pooltable be horizontal.  If the pooltable
is not horizontal, a certain symmetry is broken.  Let us call that
symmetry $S$.  If $S$ is broken, the balls tend to roll to one side, which
is no good.  The rules of pooltable physics require that $S$ be a good
symmetry.

	Similarly, the rules followed by the strong interactions
obey discrete symmetries $P$ and $CP$.  $P$ is parity and $CP$ is the product
of parity with charge conjugation.  These symmetries of the strong
interactions have been known for a long time.  In fact, the
discovery that the weak interactions violate $P$ and $CP$ was a big surprise
because physicists, used to seeing the strong interactions and also the
electromagnetic interactions obey $P$ and $CP$, had a hard time
conceiving
that these symmetries could be violated at all.

	We may imagine that the people playing snooker have done so for
a very long time.  Let us even imagine that they have always
lived on the pooltable.  They have a hard time conceiving that the
$S$ symmetry could be broken.  However, some day they discover the great
wide world.  They jump off the table and find themselves on the floor
of the playroom.  Now, to continue our analogy with the Standard Model
of particle physics, we will assume that the playroom floor is not
horizontal.  The snookerplayers are astonished to discover that the wider
world does not respect the $S$ symmetry they had become so used to.  The
playroom is skew somehow, which is very disconcerting.  But after a while
the snookerplayers become accustomed to this.  They abandon the prejudice
that $S$ should be a good symmetry.

	The snookerplayers have become comfortable with the wider world
and the fact that the $S$ symmetry is broken.  However, one of them whom
they call TSP (which could be short for Thinking Snooker Player) is
deeply troubled.  TSP realizes there is something wrong with
the world he is living in.  The playroom floor is not horizontal because
the $S$ symmetry is broken.  That's fine.  But why is the pooltable
horizontal?

	There is similarly something wrong with the Standard
Model. This is called the ``Strong $CP$ Problem''.  The Standard Model
violates $P$ and $CP$.  How can the strong interactions, which are part
of the Standard Model, conserve those symmetries?  Within the Standard
Model, it is as surprising to have the strong interactions conserve $P$
and $CP$ as it is surprising to find a horizontal pooltable in a playroom
which is itself not horizontal.  The Standard Model takes just pride in
being able to explain the violation of $CP$ in an economical and
natural way, by allowing the Yukawa couplings in the model to have
arbitrary complex phases.  This virtue is often emphasized.  However,
if the Yukawa couplings have arbitrary complex phases, then the $\theta$
angle of QCD has an arbitrary value as well, let us say any number between
zero and $2\pi$, in which case the strong interactions violate $P$ and
$CP$ in blatant fashion.  This is contrary to observation.  To be explicit,
the upper limit on the neutron electric dipole moment, which provides
the most sensitive test of $P$ and $CP$ violation by the strong interactions,
requires that $\theta$ of QCD be less than $10^{-9}$.

	His curiosity piqued, TSP sets out to check whether the pooltable
is as horizontal as it appears.  (Yeah, TSP is no casual observer.  He's
got the soul of a physicist.)  He finds that the pooltable is as horizontal
as he can make out and, after much work, having pushed to its limits the
measurement technology available to him, concludes that any deviation of
the pooltable from perfect horizontality must be characterized by an angle
less than $10^{-9}$.  Having been around and stuff, TSP knows that one
part in a billion is easier said than done.  He is astounded.  His
stomach cringes with the fear induced by the discovery of a fact at
once bizarre and unexplained.  ``Someone is playing a trick on us, that's
for sure,'' he thinks to himself.

	TSP figures the person who made the pooltable compensated for
the slant of the playroom floor by adjusting the lenghts of
the pooltable legs.  This is illustrated in Fig. 1.  To do this,
the pooltable maker measured the angle between the vertical and the
playroom floor.  The vertical direction is determined by gravity and is
manifested by the plumb, a wellknown and wonderful tool.  After having
taken his measurements, he designed the pooltable legs accordingly,
with a precision of one part per billion.  TSP muses that if the pooltable
maker has many customers, he must spend a lot of effort adjusting his
pooltables to the various angles between the vertical and the floors of
his customers' playrooms.  Each pooltable has to be individually build
to insure the $S$ symmetry that the customers demand for their snooker
playing, to the tune of one part per billion.

        Some time passes by.  One day, as TSP sat around thinking about
the life of the pooltable maker, an idea occurred to him.  If he himself
TSP were in the pooltable making business, what he TSP would do is
build each pooltable on a post that can pivot on an axle.  At the end of
the post opposite the pooltable is a big weight.  The axle is
mounted on a tripod.  TSP's contraption is illustrated in Fig. 2.  The
point is that gravity will automatically pull the weight down,
the post vertical and the pooltable horizontal.  Et voila!  You see,
all pooltables can be made identical now, with tremendous savings in
effort and production costs.  TSP gets excited at the idea of the fortune
he could make in the pooltable manufacturing business.  His pooltables
would adjust themselves automatically in any playroom,
just under the influence of gravity.  The beauty of the scheme is that
it is gravity which decides what's vertical and what is not.  So it can
do the job of making the pooltable horizontal, just by itself!

        What TSP just discovered is the analog of the Peccei-Quinn (PQ)
solution to the strong $CP$ problem of the Standard Model of particle
physics.  Peccei and Quinn slightly modified the Standard Model in
such a way as to make the $\theta$ angle of QCD a dynamical variable.
There are non-perturbative effects which produce $P$ and $CP$ violation
in QCD if the $\theta$ angle differs from zero or $\pi$.  The analog of QCD
is the physics on the pooltable; the analog of the $\theta$ angle is the
misalignment of the pooltable from the horizontal; the analog
of the non-perturbative effects that make QCD physics depend upon the
$\theta$ angle is gravity which makes pooltable physics sensitive to
lack of horizontalness of the table; the analog of $P$ and $CP$ symmetry
in QCD is $S$ symmetry in the pooltable physics; and so on.  In the
PQ mechanism, the non-perturbative effects which make QCD physics
depend upon $\theta$, pull $\theta$ to zero once the model has been
arranged so that $\theta$ becomes a dynamical variable.  In
TSP's contraption, gravity, which makes pooltable physics
sensitive to a slant of the pooltable, removes any such slant
once an axle is introduced to allow the pooltable to pivot.

        TSP is pleased with himself, although it turns out he cannot
make a fortune based on his insight.  For some reason, he is confined
to the playroom and this keeps him from going in the pooltable
manufacturing business.  More time passes by.  One day, in a more humble
mood than the one he got into following his theoretical discovery of the
mechanism that can straighten out pooltables (he had become very excited
then), a fresh idea occurs to him.  It might be that the pooltable
maker who made the pooltable where TSP lives also discovered the
mechanism for straightening pooltables and that he incorporated it
into the pooltable in TSP's room.  TSP becomes very curious about
this possibility.  Unfortunately, all around the pooltable hangs
a dark cloth which hides from view whatever supports it.  But, after
a while, TSP realizes that it is not necessary to see the support
structure to deduce whether or not the pooltable has been build
with the pooltable straightening mechanism.  The point is that
the physics of playing snooker on a pooltable with the mechanism
differs from the physics of playing snooker on a regular pooltable,
without the mechanism.  On a regular table, as in Fig. 1, when
the ball hits the rim, it bounces back with the same energy as
it had before hitting the rim.  (For the sake of argument, we
are neglecting the absorption of energy by the rubber on the rim.)
But on a pooltable which has the straightening mechanism of Fig. 2,
a ball does not bounce off the rim with the same energy because
some of the energy gets transferred to an overall oscillation of
the pooltable about its horizontal equilibrium position.  In the
past, snookerplayers always perceived that the ball bounces back
with the same energy but, of course, they had no reason to question
whether this is true with infinite precision.

	Let me digress briefly to explain what is going through
TSP's mind at this moment.  TSP's fellow snookerplayers have
always thought him a bit odd because, although TSP was recognized from
early on to be quite smart, he didn't achieve much in real life.  TSP
just sits around thinking about this and that but he never does much.
His fellow snookerplayers had thought TSP was acting very strangely
when he had insisted that there is something ``terribly wrong'' about a
horizontal pooltable in a room which is itself not horizontal.  ``What's
so wrong about that?'' they said to each other, ``It's actually
good to have a horizontal pooltable to play snooker''.  Their viewpoint
is just so, oh, totally different from TSP's.  Of course, TSP enjoys
thinking and that's why he does that rather than anything else.  So,
contrary to what his fellow snookerplayers believe, TSP has a happy life.
Still, he would like it better if he were more appreciated.  Now, with
his theoretical discovery of the pooltable straightening mechanism,
TSP sees an opportunity to impress his fellows.  If he can show
that the rules of snooker are not quite what they appear to be
and hence that there are new possibilities to the game, that is
something his fellow snookerplayers would appreciate.  They did not
care to wonder why the pooltable is horizontal even though the playroom
floor is tilted, but if balls can give up some of their energy to an
oscillation of the pooltable and hence an oscillation of the pooltable
can give extra energy to the balls, well, of course, that is very
important and they will want to know about that.

	So TSP sets hard to work.  His goal is simple.  He wants to
produce an oscillation of the pooltable and then put into evidence that
such an oscillation is occurring.  For example, he puts one ball some
place on the pooltable next to the rim.  Then he shoots another ball
very hard against the rim on the opposite side of the table.  Some
of that energy gets absorbed into an oscillation of the table.  Then
some of the energy in the table oscillation gets transferred to the
first ball which was sitting next to the table rim.  This would be
the experimental signature of the fact that the pooltable is built with
the pooltable straightening mechanism.  When TSP's fellow snookerplayers
see that energy can be transferred from one ball to another without
the balls actually touching each other, they will be astounded.  They
will want to know how this happens.  TSP will give them lectures.  TSP
will become famous.  So he hopes.

	The analog of the pooltable oscillation in case the pooltable
is built with the pooltable staightening mechanism of Fig. 2 is, of
course, an oscillation of the $\theta$ parameter of QCD if the
Standard Model has incorporated into it the Peccei-Quinn mechanism
described earlier.  The axion is the quantum of oscillation of the
$\theta$ parameter of QCD.  It is a particle in the same way that the
quantum of oscillation of the electromagnetic field, the photon, is
a particle.  To discover whether the Peccei-Quinn mechanism has
been incorporated into the Standard Model, one searches for the axion.
To search for the axion, one tries to produce a few and then detect them.
It is necessary to produce them first because they are unstable and hence
cannot be around for a long time.  (This last statement is not always true
but let's accept it for the moment.  We will return to this point
later.) To produce axions, one may take a beam of protons and dump it into
a block of material.  The axions produced may then be put into
evidence by a detector that converts their energy back into more
immediately visible forms of energy such as photons.  This experiment
and many others which try to produce and detect axions were carried
out in the late 70's and early 80's but no axions were found.  They
are the analog of the experiment TSP proposes to carry out to put
into evidence the pooltable straightening mechanism.

	As it turns out, TSP's hopes are dashed, totally, mercilessly ...
No matter how hard he tries, he does not manage to produce an oscillation
of the pooltable that is sufficiently large for him to detect.  What is
he to make of that?  What makes the pooltable horizontal to one part
in a billion if not the mechanism of Fig.2?  Must he return to the
idea that the pooltable maker adjusted the lengths of the legs with
the required precision?  At this point, TSP realizes that his ability
to put into evidence oscillations of the pooltable depends upon the
length $l$ of the lever arm between the axle and the big weigth.  If
the length $l$ is very large, it becomes very difficult to produce pooltable
oscilations by hitting balls against the pooltable rim.  TSP also notices
that when $l$ is very large, the oscillation frequency of the pooltable
is very low.  TSP now carries out detailed calculations.  He finds that
if the length $l$ is more than about three meters, his attempts to produce
and detect pooltable oscillations must fail even if the pooltable is
constructed with the pooltable straightening mechanism.  Thus, his
experiments only rule out the mechanism if $l$ is less than three
meters or, equivalently, if the oscillation frequency of the pooltable
is more than 0.18 cycles per second, which is the oscillation frequency
of a pendulum of length three meters, on Mars where TSP and his fellow
snookerplayers happen to be living.

	TSP believes he understands everything now.  The reason the
pooltable is horizontal is the mechanism of Fig. 2.  The reason he cannot
put into evidence pooltable oscillations is that the length $l$ of
the lever arm is more than three meters.   TSP thinks he ought to
be pleased with his insight, but actually he feels pretty frustrated.
He understands why the pooltable is horizontal but he cannot produce
the pooltable oscillations which would confirm his understanding and
surprise his fellow snookerplayers.  If the mechanism is implemented
with a very long lever arm $l$, there's just no way anyone will ever
be able to put into evidence pooltable oscillations.  Yet, the
mechanism works!  With some bitterness, he mutters to himself
a name for his invention.  He calls it the ``invisible pooltable
straightening mechanism'', because it works yet it can not be
visibly demonstrated.

	The analog of the ``invisible pooltable straightening mechanism''
in the world of particle physics is the PQ mechanism with
an ``invisible'' axion.  The properties of the axion depend upon a
parameter $f$, called the axion decay constant, which is analogous to
the length $l$ in the pooltable straightening mechanism.  If $f$ is very
large, then the axion becomes very light and very weakly coupled.  The
axion mass $m$ is related to the minimum oscillation frequency $\nu$ of the
$\theta$ parameter of QCD by the famous relation:  $mc^2 = h\nu$ where $h$
is Planck's constant.  So, the small mass of the axion if $f$ is large
is analogous to the low pooltable frequency if $l$ is large.  Also, the
fact that the axion is weakly coupled is analogous to the fact that
it is difficult to produce pooltable oscillations.  If $f$ is large, the
axion production and detection rates in the axion search experiments
described earlier are so low that these experiments cannot find axions
even if axions exist.  But the PQ mechanism still works!

	TSP ponders his fate.  What is the worth of theoretical insight
without experimental confirmation?  Einstein discovered General Relativity
and very soon afterwards his theory was confirmed by the measurement
of the deflection of starlight by the sun.  Democritus discovered
(correctly guessed?  What is the difference between a theoretical
discovery and a good guess?) that matter is made of atoms.  At the time
there were no experiments that could put atoms directly into evidence.
Those experiments came twenty-three centuries later ...  As he walks around
the playroom, pondering this and other questions, TSP glances for the
umptieth time at a copper plate that is affixed to the side of the
pooltable.  It reads:  `Made in Minneapolis, Minnesota, USA, home
of ``Minnesota Fats''.  TSP always wondered what is `Minnesota
Fats' ...  But he does know about the USA.  The USA is a country on Earth.

	TSP imagines how the pooltable was brought to Mars from
Earth on a spaceship.  That's actually pretty interesting because during
the trip between Earth and Mars, when the spaceship is just coasting
along, there is `no gravity'.  In that situation, the pooltable is not
oriented in any particular direction.  It would seem impossible to play
snooker then ..., but TSP is thinking about something else altogether.
What strikes him is that when the spaceship approaches Mars and prepares
for landing by firing its retrorockets, the pooltable is not
initially horizontal with respect to the direction of gravity at the place
on Mars where the spaceship is going to land. The landing on Mars is
illustrated in Fig. 3.  Only when the rockets are fired, does the big
weight of the pooltable straigthening mechanism begin to feel Mars' gravity.
It then begins to pull the pooltable horizontal with respect to the direction
of gravity on Mars, but it overshoots!  The pooltable does not get
pulled nicely to a horizontal position.  Instead, because there is no
damping mechanism, it oscillates about the horizontal.   Once it has landed,
it will oscillate about the horizontal with constant amplitude indefinitely
because it turns out that, if the length $l$ is longer than three meters,
the pooltable oscillations are so weakly coupled that they continue
for very long times, much longer than the present age of the solar system.
(We are assuming for the analogy's sake that there is no friction on the
axle about which the pooltable pivots and that the only way pooltable
oscillations get damped is by giving off energy to the large collection
of billiard balls on the table.)

        Therefore, if the pooltable in the playroom where TSP lives is
horizontal because of the so-called `invisible' pooltable straightening
mechanism, then it should be still oscillating now.  The oscillation is
a relic of the epoch when the pooltable was brought to Mars.  What is
the amplitude of this relic oscillation?  TSP realizes that the crucial
parameter is the ratio of the pooltable oscillation period to the time
scale over which Mars' gravity gets effectively turned on when the
spaceship bringing the pooltable landed on Mars.  If the landing is
very sudden as compared to the oscillation period of the pooltable,
for example if the spaceship is in free fall towards the Martian
surface till the very last millisecond when the retrorockets are
turned on full blast to decelerate the spaceship and bring it to
zero velocity just before it lands, then the final amplitude of
the oscillation is the initial misalignment angle which is a random angle
between zero and 180 degrees and which we have no reason to assume to be
particularly small.  This possibility is incompatible with the present
state of the pooltable because the pooltable does not appear to oscillate
at all now.  If, on the other hand, the landing occurs very slowly, i.e. if
the retrorockets are fired long before the spaceship lands and therefore
Mars' gravity is turned on very progressively, then the amplitude of
oscillation decreases while the landing occurs.  The switch-on of gravity
is adiabatic in this case, and the oscillation amplitude decreases as
the inverse of the square root of the oscillation frequency, and the latter
increases as the square-root of Mars' apparent gravity.

	TSP carries out careful observations on the pooltable to determine
whether it is oscillating at present since he realizes now that a
relic oscillation is the telltale sign of the pooltable straightening
mechanism.  He does not detect any and places an upper limit of $10^{-12}$
on the present oscillation amplitude of the pooltable.  This rules out
the possibility of making the pooltable straightening mechanism invisible
at will by lengthening the lever arm $l$ because the longer $l$, the lower the
oscillation frequency of the pooltable and, comparatively, the more sudden
the switch-on of gravity when the spaceship bringing the pooltable landed
on Mars, and hence the larger the amplitude of relic pooltable oscillations.
{}From a NASA publication that happens to be laying around in the playroom,
TSP can deduce the time scale over which the retrorockets are fired when
landing on Mars which is also the time scale over which Mars' gravity gets
effectively turned on.  From this he can figure out the amplitude of
relic pooltable oscillations as a function of $l$.  He finds that the
upper limit of $10^{-12}$ on relic pooltable oscillations requires that
$l$ be smaller than 10 meters.  TSP is very excited about this result.
On the one hand, $l$ must be larger than 3 meters because he was unable
to produce and detect pooltable oscillations.  On the other hand, $l$
must be smaller than 10 meters because he was unable to detect
relic pooltable oscillations.  It seems TSP is closing in on a
resolution of the horizontal pooltable mystery.

	The switch-on of gravity when the spaceship approaches
Mars is analogous to the switch-on of non-perturbative QCD effects
when the universe is about $10^{-7}$ seconds old and the temperature is
about one GeV.  The relic pooltable oscillations are analogous to the
coherent axion field oscillations that constitute the present axion
cosmological axion energy density if $f$ is large.  The requirement that
the axion cosmological energy density not overclose the universe puts
an upper limit on $f$ and hence a lower limit on the axion mass.  Just as
TSP found lower and upper limits on the length $l$ of the lever arm in
the pooltable straightening mechanism, there are lower and upper limits
on the axion decay constant $f$.  If the axion mass is near its lower
limit, axions may be the dark matter of the universe.  Experiments
presently under way attempt to detect the axion field oscillations
that constitute the dark matter in our galaxy.

	TSP figures out a means of detecting relic pooltable oscillations
if $l$ is in the range 3 to 10 meters, or equivalently, if the frequency
of relic pooltable oscillations is in the range 0.18 to 0.097 cycles
per second.  His device is just a simple high quality oscillator placed
on the pooltable.  See Fig. 4.  The oscillator frequency is tunable by
changing the mass at the end of the spring.  TSP plans to slowly change
the frequency.  When it equals that of relic pooltable oscillations,
his oscillator will get excited.  TSP should be able to see this effect.

        Will TSP succeed in his latest venture and solve at last the
mystery of the horizontal pooltable?  We don't know yet but if he
succeeds, there may be a sequel to this story.
\bigskip
\bigskip
\bigskip
\noindent {\bf Acknowledgements:}
\bigskip
I would like to thank Jim Ipser and Charles Thorn for reading the
manuscript, and Cynthia Chennault for stylistic suggestions.  This work
was supported in part by the US Department of Energy under contract
\#DE-FG05-86ER40272.
\bigskip
\bigskip
\bigskip
\noindent {\bf Figure Captions:}
\bigskip
\item {} Fig. 1.  A horizontal pooltable on a slanted floor.
\item {} Fig. 2.  The pooltable straightening mechanism.
\item {} Fig. 3.  The pooltable in the spaceship about to land on Mars.
\item {} Fig. 4.  A detector of relic pooltable oscillations.
\bigskip
\bigskip
\bigskip
\noindent {\bf Bibliography:}
\bigskip
The Strong CP Problem and axion physics have been reviewed in:

\item {\bf -} J. E. Kim, Phys. Rep. \underbar {150} (1987) 1;
\item {\bf -} H.-Y. Cheng, Phys. Rep. \underbar {158} (1988) 1;
\item {\bf -} R. D. Peccei, in ``CP Violation'', ed. by C. Jarlskog,
World Scientific. Publ., 1989, pp. 503-551.
\item {\bf -} M. S. Turner, Phys. Rep. \underbar {197} (1990) 67.

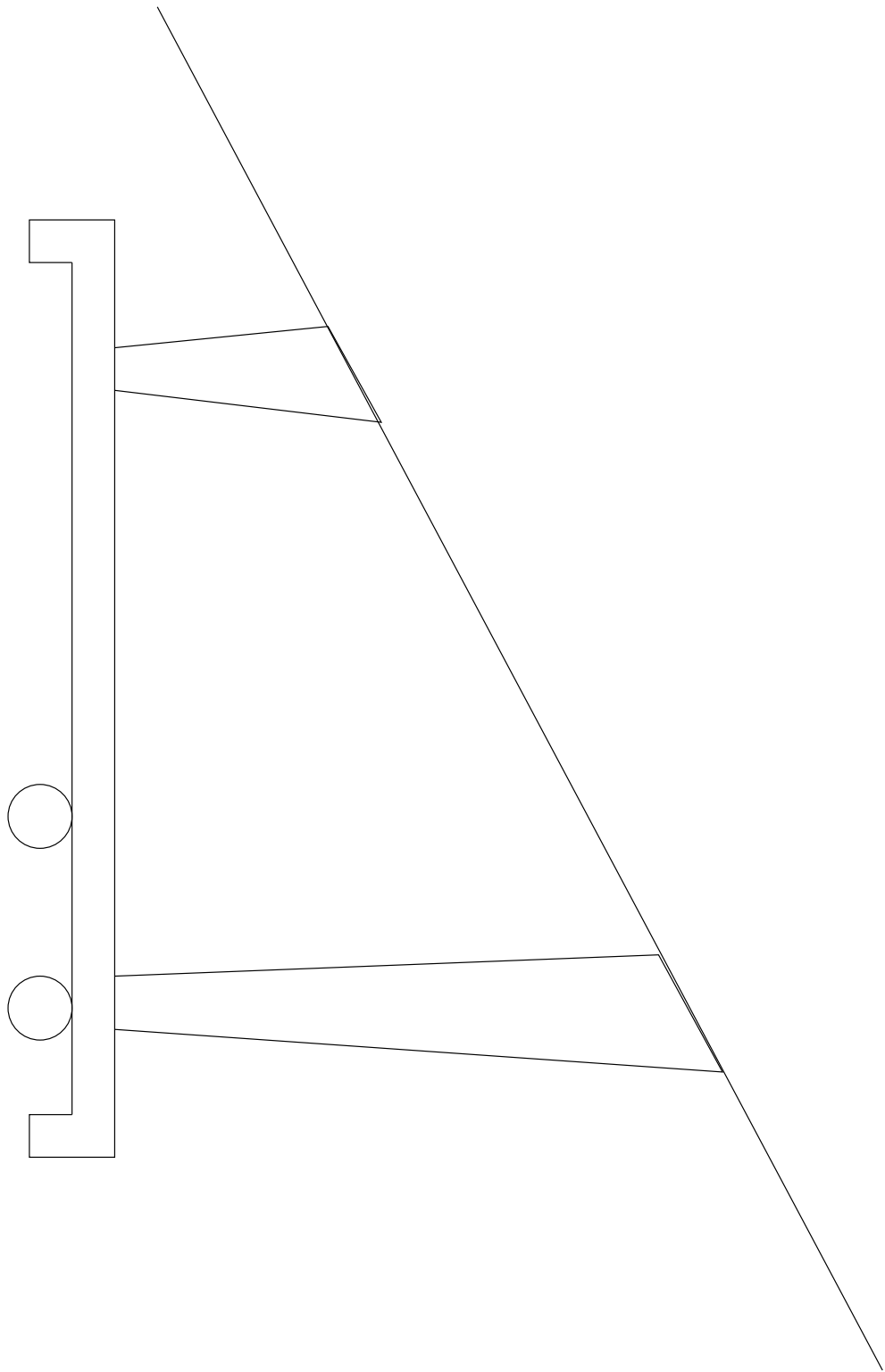
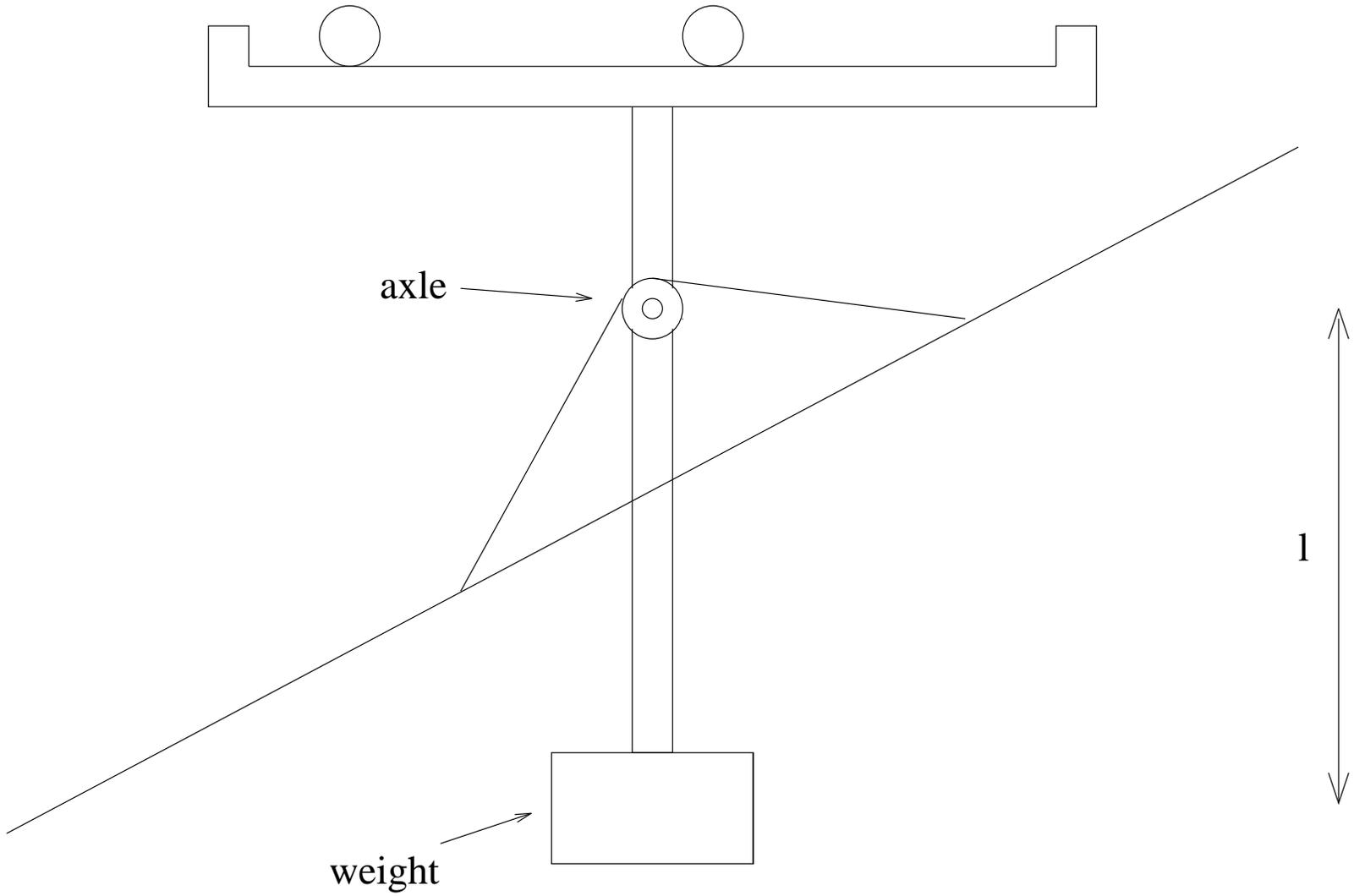
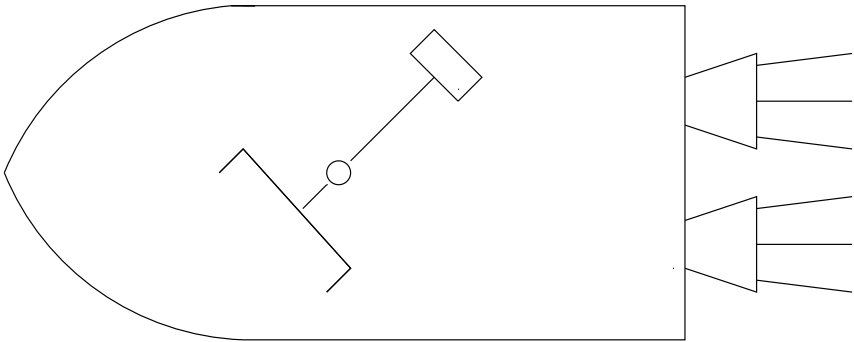
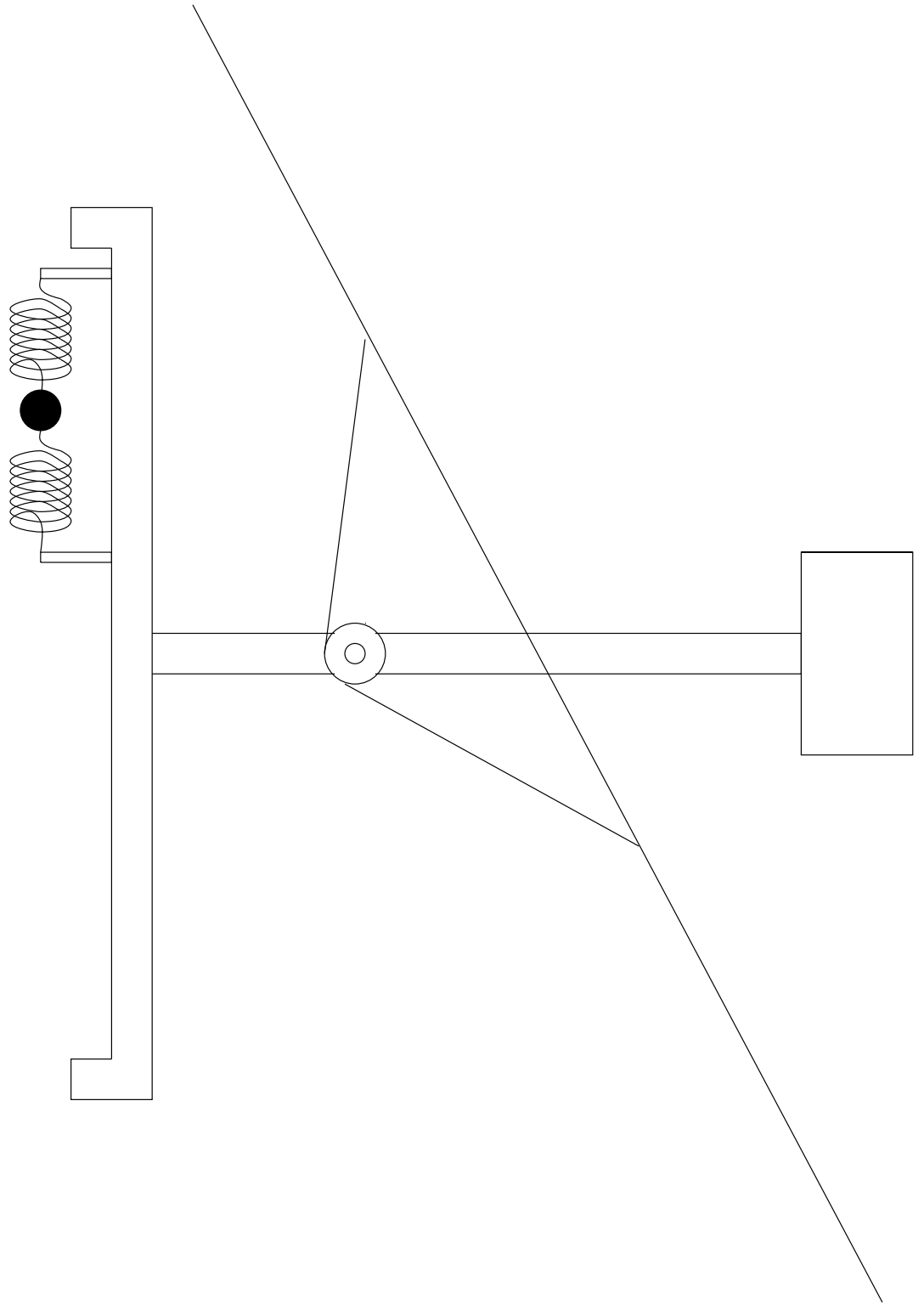
\end